\documentclass[%
 reprint,
 amsmath,amssymb,
 aps,
prl,
]{revtex4-2}
\usepackage{CJK}
\usepackage{graphicx}
\usepackage{dcolumn}
\usepackage{bm}
\usepackage{xcolor}

\usepackage{siunitx}
\usepackage{xspace}
\newcommand{\um}{\si{\micro m}}

\begin{document}
\begin{CJK*}{UTF8}{gbsn} 
\preprint{APS/123-QED}

\title{Low-power threshold optical bistability enabled by hydrodynamic Kerr nonlinearity of free-carriers in heavily doped semiconductors}

\author{Huatian Hu (胡华天)}
 \email{huatian.hu@iit.it}
\author{Gonzalo \'Alvarez-P\'erez}
\author{Tadele Orbula Otomalo}
\author{Cristian Cirac\`i}%
 \email{cristian.ciraci@iit.it}
\affiliation{%
 Istituto Italiano di Tecnologia, Center for Biomolecular Nanotechnologies, Via Barsanti 14, 73010 Arnesano, Italy
}%

\date{\today}

\begin{abstract}
We demonstrate nanoscale optical bistability at an exceptionally low power threshold of 1 mW by leveraging Kerr-type hydrodynamic nonlinearities due to the heavily doped semiconductor's free carriers. This high nonlinearity is enabled by a strong coupling between metallic nanopatch surface plasmons and longitudinal bulk plasmons (LBP) that arise in heavily doped semiconductors due to the nonlocality. Through the coupling, an efficient, near-unity conversion of far-field energy into LBP states could be achieved. These findings offer a viable approach to experimentally probe LBPs and lay the groundwork for developing efficient and ultrafast all-optical nonlinear devices.

\end{abstract}

\maketitle
\end{CJK*}
Attaining efficient, compact, and ultrafast photon-photon interactions via the nonlinear optical Kerr effect has long been sought-after due to their potential applications in all-optical information processing and storage \cite{boyd2008nonlinear,krasavin2018free,almeida2004optical,gibbs1982room,un2023electronic,cox2014electrically,christensen2015kerr,cox2016quantum,ryabov2022nonlinear,niu2023all}.
The advent of artificial intelligence in recent years has led to an exponential increase in demand for low-power computation solutions, bringing a renewed interest in integrated all-optical systems \cite{feldmann2019all}.

Semiconductors represent in this context an interesting class of materials, as they are compatible with large-scale fabrication methods for integrated devices while also supporting some of the largest optical nonlinearities \cite{lee2014giant,taliercio2019semiconductor}. 
When heavily doped, semiconductors can undergo a transition from the quantum regime, where their response is governed by single-particle excitations, to the classical regime of collective plasmon oscillations \cite{taliercio2019semiconductor}. 
Due to the hydrodynamic nature of non-equilibrium carriers, collective plasmons can exhibit an intrinsic and strong free-carrier optical nonlinearity \cite{scalora2010second,ciraci2012second,krasavin2018free,khalid2020enhancing,de2021free,de2022impact}.
Although these properties occur, in principle, in all degenerate electron systems, they are significantly larger for materials with low equilibrium density and small effective mass, such as doped InP and InGaAs \cite{de2022material}.
In fact, very recently, it has been experimentally shown by measuring third-harmonic generation from heavily doped InGaAs nanoantennas that free-electron nonlinearities can easily overcome intrinsic $\chi^{(3)}$ nonlinearities \cite{rossetti2024origin}. 
The drawback however is that free-electrons inside a material mostly respond to external optical fields that oscillate below the plasma frequency, i.e., in the metallic regime as surface plasmons.
This confines hydrodynamic nonlinear interactions at the material surface\cite{de2022impact,ciraci2012second,scalora2010second}, severely limiting the Kerr effect.

In this context, longitudinal bulk plasmons (LBPs) -- charge-density waves occurring in the \textit{bulk} of spatially dispersive materials \cite{ruppin_optical_1973,ruppin2001extinction,raza2011unusual} -- become very relevant. 
The relative distribution of the electrons over a positively charged background generates their characteristic longitudinal electric fields \cite{benisty2022introduction}.
LBPs have been observed in thin metal films where they naturally occur in the ultraviolet energy range \cite{anderegg_optically_1971,lindau_experimental_1971}.
More recently, LBP resonances have been observed in transparent conducting oxide materials in the infrared, such as doped cadmium oxide \cite{de2018viscoelastic}.
Existing above the plasma frequency, the LBP sacrifices the typical field confinement associated to surface plasmons for a much larger active volume for nonlinear interactions, unlocking a new mechanism for low-power Kerr nonlinearity at the nanoscale.
However, because of their longitudinal nature and their large wavevectors, LBPs can only be excited very weakly from the far-field \cite{christensen2014nonlocal}, conventionally rendering their use in applications very difficult. 

From a theoretical perspective, LBP modes can be excited in finite-size systems only if the response of the material is \textit{nonlocal} \cite{ruppin2001extinction,christensen2014nonlocal}.
One way to account for nonlocality in degenerate systems is to model the classical free-electron dynamics incorporated with density-dependent energy potentials that encapsulate their quantum properties.
This approach is usually referred to as \textit{hydrodynamic theory} (HT) \cite{ciraci2016quantum, toscano2015resonance}.
For a proper choice of the kinetic energy potential, it can be shown that the HT is equivalent to the Schr\"odinger equation for single particle systems (or more precisely, single orbital systems) \cite{ciraci2017current}.
Interestingly, the HT maintains its \textit{quantum} structure for systems characterized by many electrons, as long as they can be considered as a collective quasi-particle, such as plasmons.
With this premise, quantum-size resonances in quantum-confined systems (i.e., quantum wells and quantum dots) in traditional semiconductors can be seen from a theoretical point of view as equivalent to quantized LBP  modes in heavily doped semiconductor nanostructures, with the difference that instead of single-electron states, the HT accounts for many-particle states in which all electrons are coupled via electron-electron and electron-photon interactions.
Quantum-well states have been indeed experimentally observed in metals \cite{silberberg1992optical} and have been associated with the enhancement of Kerr nonlinearity \cite{qian2016giant}, although the underlying mechanism is not fully understood.

In this letter, we consider for the first time the hydrodynamic contributions to Kerr nonlinearity in heavily doped semiconductors and demonstrate that LBP resonances could indeed drive a very large Kerr effect in subwavelength volumes.
Moreover, by strongly coupling heavily doped semiconductors with traditional plasmonic systems, we show that it is possible to drive as much as near-unity (97\%) of the far-field energy into the LBP modes.
As a result, an ultralow-threshold ($\sim 1.55$ mW/\um) and compact optically bistable device is proposed.

The many-body nature of a free-electron gas driven by electromagnetic fields $\mathbf{E}(\mathbf{r},t)$ and $\mathbf{H}(\mathbf{r},t)$, can be described, within the HT, through two macroscopic fields, the electron density $n(\mathbf{r},t)$ and velocity $\mathbf{v}(\mathbf{r},t)$:

\begin{equation}\label{eq:HT-EqMotion}
\begin{aligned}
m_e\left( {\frac{\partial }{{\partial t}} + \mathbf{v} \cdot \nabla  + \gamma } \right)\mathbf{v} = -e({\bf{E}} + \mathbf{v} \times {\mu_0 }{\bf{H}}) - \nabla \frac{{\delta {G}[n]}}{{\delta n}},
\end{aligned}
\end{equation}
with $m_e$ and $e$ being the effective electron mass and the electron charge, respectively; $\gamma$ is the phenomenological damping rate and $\mu_0$ is the vacuum permeability.
The terms from the left to right in Eq.\eqref{eq:HT-EqMotion} account for time-evolution, convection, dissipation, Coulomb, Lorentz, and internal energy potential, i.e. quantum pressure, respectively.
Since we are not interested in electron spill-out effects, here we consider the energy functional within the Thomas-Fermi (TF) approximation $G[n]\simeq T_\mathrm{TF}[n]$ \cite{ciraci2016quantum}.
Introducing the polarization field defined as $\partial{\mathbf{P}}/\partial t=\mathbf{J}=-en\mathbf{v}$, expanding the nonlinear terms up to the third order and neglecting the higher-order nonlinear terms \cite{de2021free}, we obtain the following equation (see Supplemental Material \cite{supp} for detailed derivation):
\begin{eqnarray}\label{eq:constitutive}
\frac{{{\partial ^2}{{\bf{P}}}}}{{\partial {{t}^2}}} + \gamma \frac{{\partial {{\bf{P}}}}}{{\partial t}} = \frac{{{n_0}{e^2}}}{m_e}{{\bf{E}}} +  \beta ^2\nabla (\nabla  \cdot {{\bf{P}}}) + {{\bf{S}}^{\rm{NL}}},
\end{eqnarray}
where the first-order quantum pressure term has been written as $\nabla(\frac{\delta {T}_\mathrm{TF}}{\delta n})_1 = \frac{m_e}{en_0}\beta ^2\nabla (\nabla  \cdot {{\bf{P}}})$, where $\beta^2=\frac{10}{9}\frac{c_{\rm{TF}}}{m_e}n_0^{2/3}$, $c_{\mathrm{TF}}=\frac{\hbar}{m_e^2}\frac{3}{10}(3\pi^2)^{2/3}$.
All the higher-order terms containing the nonlinear responses to the external field and electron-electron interactions are encapsulated into the total nonlinear source $\mathbf{S}^{\rm NL}$, which will be elaborated on later in Eq.\eqref{eq:nonlinearsources}.
Its origin should not be confused with the bulk nonlinear susceptibility $\chi^{(3)}$ that comes from the anharmonic potential experienced by the electrons in crystals.
The former is mostly a nonlocal response depending on the gradient of the fields, the latter is a local response that will be considered separately in the wave equation as Eq.\eqref{eq:waveequation}.

\begin{figure}[t]
\includegraphics[width=0.48\textwidth]{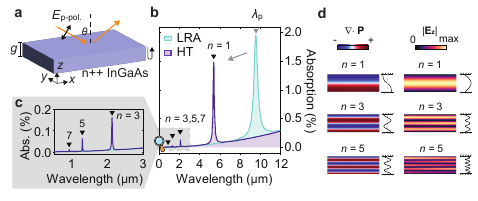}
\caption{\label{fig:1} Optical absorption of a heavily doped InGaAs thin slab. (a) Schematic. (b) Linear absorption spectra of the thin slab with local-response approximation model (LRA, cyan) and nonlocal hydrodynamic theory (HT, purple). (c) Zoom-in plot of the higher order LBPs (d) The induced charge $\nabla \cdot \bf P$ and electric field of LBPs with an order of $n=$ 1, 3, 5.}
\end{figure}

Let us start with the simplest system supporting LBP modes: a thin slab excited by $p$-polarized light at oblique incidence of $\theta=60$ degrees, as depicted in Fig.\ref{fig:1}a. An electric field with a non-zero normal component at the material interface is required to align with the charge oscillations for driving LBP.
The linear properties of LBP can be obtained in the frequency domain by solving the linear part of Eq.\eqref{eq:constitutive} (i.e., $\mathbf{S}^{\rm NL}=0$) coupled to the wave equation: 
\begin{equation}\label{eq:MaxwellFreq}
\begin{aligned}
\nabla \times \nabla \times \mathbf {E} - \varepsilon_{\mathrm{r}}\frac{\omega^2 }{c^2} \mathbf{E} - \mu_0\omega^2 \mathbf{P}=0,
\end{aligned}
\end{equation}
where $\varepsilon_{\rm r}$ is the relative permittivity accounting for the dielectric local response and $c$ is the speed of light in vacuum.

We consider a thin layer ($g=10$ nm) of heavily doped InGaAs with equilibrium carrier density $n_0=6\times10^{18}$ cm$^{-3}$.
The doped InGaAs can be described as a Drude-like material with $\varepsilon_{\mathrm{inf}}=12$, $\gamma=8.9$~ps$^{-1}$, $m_e=0.041m_0$ ($m_0$ being the electron mass) \cite{rossetti2024origin}, such that the screened bulk plasma wavelength lays at $\lambda_{\rm{p}}= 2\pi c \sqrt{m_e\varepsilon_0\varepsilon_{\mathrm{inf}}/n_0e^2}= 9.56$~\um.

The linear absorption spectra with and without quantum pressure are shown in Figs.\ref{fig:1}b,c.
With local-response approximation (LRA), a prominent absorption peak arises due to the epsilon-near-zero (ENZ) enhancement at $\lambda_{\rm p}$. However, as electron-electron interactions are considered with the HT, the ENZ resonance exhibits a noticeable blueshift, accompanied by the emergence of additional higher-order resonances fulfilling the LBP condition of $\varepsilon(\omega,\mathbf{k})=0$.
Only LBP modes with odd orders ($n=1, 3, 5, \ldots$) have a net dipole moment that can be excited.
In Fig.\ref{fig:1}d, the charge density and electric field distributions clearly show the nodes and antinodes of the standing waves formed by the charge-density oscillations. 

Although appreciable, LBP modes in a thin slab display an absorption, in the best case, of only $\sim$1.5\%, (similar values are observed for the reflectance).
Higher-order resonances, shown in Fig.\ref{fig:1}c, have even smaller absorption peaks (less than 0.2\%).  
This is not surprising since the LBPs can only couple weakly to the transverse plane wave in free space. This characteristic hinders both the experimental observation as well as the application in linear and nonlinear devices.

\begin{figure}[b]
\includegraphics[width=0.48\textwidth]{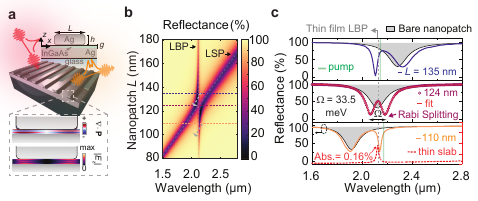}
\caption{\label{fig:2} Hybrid bulk-plasmon slab-nanopatch system.  (a) Schematic of the hybrid structure; the height $h$ and periodicity $P_x$ of the antenna are 50 nm and 200 nm, respectively. The length $L$ can vary. Bottom panels: the induced charges and electric field of the LBP in the thin spacer.  (b) Strong coupling between LSP and LBP with a clear Rabi splitting in reflectance spectra, where three particular cases of different $L$ marked by dashed lines are shown in (c) for details. The grey shades in (c) represent the reflectance of the bare nanopatch without nonlocality. The reflectance spectrum of the on-resonance case ($L=124$ nm) is fitted with a Rabi splitting of $\Omega=33.5$ meV. Green shades mark the center wavelengths of the pump for probing Kerr nonlinearities.}
\end{figure}

To overcome this issue, we consider a hybrid system constituted by a thin layer of heavily doped InGaAs sandwiched between a silver film and an array of silver stripes (Fig.\ref{fig:2}a), which is experimentally feasible \cite{jeannin2023low}.
Such systems are usually referred to as nanopatch antennas, in analogy with their radio-frequency homologue.
Nanopatch antennas have an absorption cross-section that is much larger than their physical dimensions and can concentrate far-field radiation into deep subwavelength nanogaps \cite{baumberg2019extreme}.
Additionally, the strong nanogap normal fields (${\bf E}_z$) are perfectly aligned with the longitudinal fields of the LBP supported by the semiconductor film (lower panel in Fig.\ref{fig:2}a).

The hybrid structure is designed as follows:
since the LBP is mostly determined by the thickness and doping of the InGaAs, we fix $g$ and $n_0$ as the same as those described in Fig.\ref{fig:1} such that the third-order LBP resonance is within the range of the nanopatch fundamental mode, i.e. $\sim 2.15$~\um, as shown in Fig.\ref{fig:2}b.
By adjusting the length $L$ of the nanopatch antenna, we can tune the resonances of the antenna localized surface plasmon (LSP) and its spectral overlap with the LBP.
Because of the relatively large size and high equilibrium electron density of the nanopatches, the nonlocality and free-electron nonlinearity can be negligible. Thus, we treat silver as a local Drude metal with $\varepsilon_{\mathrm{inf}}=4.039$, $\omega_\mathrm{p}=9.172$ eV, $\gamma=0.0207$ eV \cite{johnson1972optical}.
 
The detuning between the LBP and LSP is an essential factor for the coupling.
In Fig. \ref{fig:2}b, we map the reflectance of the system as a function of the incident wavelength and nanopatch $L$, which highlights the anti-crossing dispersion between the LBP and LSP. 
The  map shows anti-peaks down to virtually zero reflectance, unveiling a nearly perfect absorption behavior \cite{moreau2012controlled}.
Even when the two modes are quite detuned (e.g., $L=140$ nm), the absorption (81.7\%) of the LBP is still enormously enhanced compared to the bare InGaAs slabs (0.16\% in Fig.\ref{fig:1}c). This is due to the broad linewidth of the LSP resonances of the nanopatch, which can still provide decent field enhancements at certain detunings.
This is a substantial first result as the hybrid mode provides a practical and efficient way to access the elusive LBP. 

When the two uncoupled states perfectly overlap, i.e. for $L=124$~nm, they hybridize into two modes with identical linewidth.
By fitting the Rabi splitting with a temporal coupled-mode theory \cite{hu2021unified}, it is possible to confirm that the system is working in a strong coupling regime.
We have found that the relation between the coupling strength $g_c$ (contributing to Rabi splitting $\Omega=2g_c=33.5$~meV) and the sum of half-width at half-maximum of each bare mode $\Sigma_i(\Gamma_i/2)$ satisfies the well-known condition $4g_c/\Sigma_i(\Gamma_i)=1.26>1$, which indicates by a large margin that the system is indeed working in the strong coupling regime.
As a consequence, the two resulting dressed states (i.e., the two peaks in the spectra) are hybrid bulk-surface plasmons.
These hybrid modes hold the potential to inherit both the strong field enhancement from the LSP and strong free-electron nonlinearity from the LBP, providing a promising direction for the development of efficient and low-threshold nonlinear devices. 

To prove so, let us investigate the hydrodynamic Kerr nonlinearity supported by the dressed LBP state in the hybrid system. The free-carrier Kerr-type nonlinearity in the doped semiconductor is assumed to mainly come from the third-order hydrodynamic contributions in the constitutive relation Eq.\eqref{eq:constitutive}, which arise from the convection and TF pressure, i.e. $\mathbf{S}^{\mathrm{NL}}=\mathbf{S}^{\mathrm{cv}}+\mathbf{S}^{\mathrm{TF}}$, where (see \cite{supp} for derivations):
\begin{subequations}
\label{eq:nonlinearsources}
\begin{equation}   
 \begin{aligned}
\label{eq:hydroKerr}
{{\bf{S}}^{\rm{cv}}} = &-\frac{1}{{{e^2}n_0^2}}\Big[ (\nabla \cdot {{\bf{P}}}){{{\bf{\dot P}}}}(\nabla \cdot {\bf{\dot P}}^*) + (\nabla \cdot {{\bf{P}}}){\bf{\dot P}}^*(\nabla \cdot {{{\bf{\dot P}}}}) \\
&+ (\nabla \cdot {\bf{P}}^*){{{\bf{\dot P}}}}(\nabla \cdot {{{\bf{\dot P}}}}) + \left( {\nabla \cdot {{\bf{P}}}} \right){{{\bf{\dot P}}}} \cdot \nabla {\bf{\dot P}}^* \\
&+ \left( {\nabla \cdot {{\bf{P}}}} \right){\bf{\dot P}}^* \cdot \nabla {{{\bf{\dot P}}}} + \left( {\nabla \cdot {\bf{P}}^*} \right){{{\bf{\dot P}}}} \cdot \nabla {{{\bf{\dot P}}}} \\
&+ {{{\bf{\dot P}}}} \cdot {{{\bf{\dot P}}}}\nabla \left( {\nabla \cdot {\bf{P}}^*} \right) + 2{{{\bf{\dot P}}}} \cdot {\bf{\dot P}}^*\nabla \left( {\nabla \cdot {{\bf{P}}}} \right)\Big]
\end{aligned}
\end{equation}
\begin{equation}\label{eq:hydroKerrTF}
\begin{aligned}
{{\bf{S}}^{\rm{TF}}} = - \frac{1}{{9}}\frac{{{\beta ^2}}}{{{e^2}n_0^2}}\nabla \left[ {{{\left( {\nabla \cdot {{\bf{P}}}} \right)}^2}\left( {\nabla \cdot {\bf{P}}^*} \right)} \right]
\end{aligned}
\end{equation}
\end{subequations}
The ``$\cdot$" over the variables denotes the time derivative, while the ``$*$" denotes the complex conjugated. 
Note that because $\mathbf{S}^\mathrm{NL}\propto 1/n_0^2$, these nonlinear terms, usually neglected in noble metals, become significant in semiconductors \cite{de2022material}. In our calculation, the equilibrium density $n_0$ in heavily doped InGaAs is nearly 4 orders of magnitude less than that of noble metals, enhancing nonlinear sources by 8 orders of magnitude.

Equations~\eqref{eq:constitutive} and \eqref{eq:nonlinearsources} are solved with the wave equation for the magnetic vector potential $\mathbf{A}(\mathbf{r},t)$:
\begin{eqnarray}\label{eq:waveequation}
\nabla \times \nabla \times \mathbf{A}+  \frac{\varepsilon_{\mathrm{r}}}{c^2}\frac{\partial^2 \mathbf{A}}{\partial t^2}+\mu_0\frac{\partial }{\partial t}(\mathbf{P}+\mathbf{P}^{\mathrm{NL}}_{\mathrm{d}})=0,
\end{eqnarray}
The vector potential $\bf A$ works in the Coulomb gauge with a vanishing gradient of the scalar potential due to the absence of a static field, such that the electric field $\mathbf{E}= -\partial \mathbf{A}/\partial t$ and magnetic field $\mu_0 \mathbf{H}=\nabla \times \mathbf{A}$.
The linear and nonlinear hydrodynamic contributions are coupled into the wave equation through a polarization vector $\bf P$.
The lattice Kerr nonlinearity from the dielectric $\chi^{(3)}=1.6\times 10^{-18}$ m$^2/$V$^2$ is considered through another polarization term $\mathbf{P}^{\mathrm{NL}}_{\mathrm{d}}=3\varepsilon_0\chi^{(3)}|{\bf E}|^2{\bf E}$ \cite{boyd2008nonlinear}.  
The contributions from the free carriers (i.e., ${\bf S}^{\rm NL}$) and the dielectric  $\chi^{(3)}$ (i.e., $\mathbf{P}^{\mathrm{NL}}_{\mathrm{d}}$) will be separated and compared \cite{supp}.
Other nonlinear contributions including higher-order nonlinear process, cascaded process are beyond the scope of this letter and have been excluded from consideration.

The system of Eqs.\eqref{eq:constitutive},\eqref{eq:nonlinearsources} and \eqref{eq:waveequation} is solved with a finite-element method through a customized implementation carried out using COMSOL Multiphysics.
To reduce the computational cost, we have implemented a time-domain pulse-envelope method in which the fast time-varying components from the full time-evolving field are removed, i.e. all the fields are expressed as $\mathbf{F}(\mathbf{r},t)=\widetilde{\mathbf{F}}(\mathbf{r},t)e^{-i\mathbf{k}\cdot\mathbf{r}+i\omega t}$, with ${\bf F} ={\bf E}, \mathbf{H},\mathbf{P}, {\bf A}$. 
Here, $\mathbf{k}$ and $\omega$ are the carrier wave vector and frequency, respectively, of the initial pulse.
The tilde indicates the corresponding slowly varying envelope field.
Solving for the envelopes allows for a much larger time step (e.g., $\sim 10^2$ times of the wave period).
To take into account the transformation above, all the equations must be modified assuming the following transformations: $\nabla  \to \nabla  - i{\bf{k}}$ and $\partial /\partial t \to \partial /\partial t + i\omega$. 
In our calculations, we neglect the hydrodynamic nonlinear contribution from silver due to its high equilibrium charge density, which leads to a negligible nonlinear contribution \cite{de2021free}.

To understand the role played by the LBP-LSP coupling in the nonlinear process, we now study in detail three specific detuning situations, summarized in Fig.\ref{fig:2}c.
As the LBP is red-detuned from the LSP ($L=110$ nm, bottom panel of Fig.\ref{fig:2}c), we use the sharp resonance at 2.142 $\mu$m for achieving optical bistability.
We set the pump pulse carrier wavelength slightly above the resonance, at 2.165 \um, to probe the hysteresis loop. As a result, with the increase of pumping power, the Kerr nonlinearity will shift the resonance toward lower energies so that the resonance will gradually overlap with the pump frequency.
As shown by the guide-to-eyes grey arrow in Fig.\ref{fig:2}b for $L=110$ nm (orange line), the reflectance would see a decrease until it crosses the dip and then experiences an increase again.
\begin{figure}[t]
\includegraphics[width=0.45\textwidth]{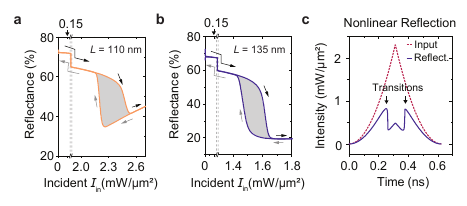}
\caption{\label{fig:3} Hysteresis loop of two-dimensional nonlinear nanopatch systems.  Prominent optical bistabilities with two particular antenna sizes: (a) $L=110$~nm, (b) $L=135$~nm. (c) Nonlinear reflection (pulse distortion) due to the bistability with $L=135$~nm.}
\end{figure}
By modulating the input intensity, we can generate the full hysteresis loop shown in Fig.\ref{fig:3}a, whose reflectance has bistable states with ``on" and ``off" states between $\sim$60\% to 30\% with a threshold starting around 2.45 mW/\um$^2$.
This situation could have a higher reflectance at the starting point but the red-detuned LBP tends to move away from LSP which degrades the low reflectance dip. Likewise, we could switch off the hydrodynamic nonlinear sources ($\mathbf{S}^{\rm NL}=0$) and study the responses solely from the enhanced dielectric nonlinearity $\mathbf{P}^{\mathrm{NL}}$.
The results show no bistability, confirming that the hydrodynamic Kerr nonlinearity is dominating \cite{supp}.

While the situation in which the LBP and LSP resonances are overlapped (Fig.\ref{fig:2}c, with $L=124$ nm) is interesting from a linear response point of view, it does not represent a viable starting point for carrying out bistability since the LBP sharp resonance is hybridized and hence broadened.
This results in a higher intensity threshold and narrower hysteresis loop.

Let us now consider the opposite situation where the LBP is blue-detuned from the LSP, taking $L=135$ nm for instance, and tuning the pumping wavelength at 2.135 \um.
Instead of getting away from the LSP, the LBP is now pushed by the nonlinearity towards the LSP, as shown by the guide-to-eyes grey arrow in Fig.\ref{fig:2}b for $L=135$ nm (purple line).
The reflectance will decrease and cross the dip, instead of increasing as in the previous situation, due to the broadening of the branch in Rabi-splitting.
This causes a ``flat" off-state in Fig.\ref{fig:3}b with 20\% reflectance.
We can interpret this result as a self-modulated dynamic transition from detuned coupling to resonant strong coupling driven by the nonlocal Kerr nonlinearity.
Moreover, the LBP is moving towards LSP to gradually get a larger field enhancement and greater coupling due to the spectra overlap. 
This further lowers the intensity threshold for achieving optical bistability: 1.55 mW/\um$^2$ for achieving 70\% to 20\% branches with 5.4 dB modulation depth. Again, the dielectric $\chi^{(3)}$ contribution is negligible compared with the hydrodynamic ones \cite{supp}, revealing the advantage of the free carriers as a nonlinear source for the Kerr effect.

In Fig.\ref{fig:3}c we depict the nonlinear reflection against the input pulse.
We can observe that the reflected pulse is dramatically distorted due to the transitions at a threshold intensity.
Although this threshold is comparable to previously considered systems \cite{huang2015optical,almeida2004optical,gibbs1982room,christensen2015kerr,argyropoulos2012boosting}, the required intensity could be further reduced to $\sim\mu$W/\um$^2$ if one can overcome the high computational cost and consider a full three-dimensional nanopatch antenna system, which would provide an extra enhancement due to the additional transverse confinement.  

In conclusion, we proposed a scheme to observe and viably use the free-carrier LBP of heavily doped semiconductors in nonlinear optical applications.
By employing a hybrid system, we theoretically demonstrated that it is possible to leverage the strong coupling between LBP and LSP to detect the presence of LBP. 
More importantly, based on the strong coupling, we explored the impact of the hydrodynamic Kerr nonlinearity, demonstrating large contrast bistability with a power threshold of $\sim 1$ mW (considering a full patch and a focused beam area to the diffraction limit). 
Our findings open a viable avenue for ultrafast all-optical integrated circuits.

We acknowledge funding from the European Innovation Council through its Horizon Europe program with Grant Agreement No. 101046329.
Views and opinions expressed are those of the authors only and do not necessarily reflect those of the European Union. Neither the European Union nor the granting authority can be held responsible for them.

\nocite{*}

\bibliography{apssamp}

\end{document}